# Exploring Two-Dimensional Empty Space

A. K. Geim

National Graphene Institute, University of Manchester, Manchester M13 9PL, United Kingdom

*Research on graphene is proving to have more lives than a cat, repeatedly coming back in new incarnations including graphene's recent alter ego, two-dimensional emptiness.*

The advent of graphene and other two-dimensional (2D) crystals[1] has given rise to one of the most active research fields in condensed matter physics and materials science. While the properties of graphene per se had largely been explored within the first few years after its isolation, the field has not withered away but instead continued to expand, with new topics emerging seemingly from nowhere every couple of years. After graphene, it was the turn of other atomically thin materials like 2D hexagonal boron nitride (hBN), various chalcogenides, black phosphorous, etc. which provided interesting results and additional excitement. Then the development of van der Waals (vdW) assembly offered practically infinite possibilities to combine 2D crystals in designer heterostructures[2]. One of many consequences of the vdW technology was a possibility to study 2D materials poorly stable in air and inaccessible experimentally before. This gave yet another boost to the ever-expanding field. Even graphene itself experienced a true reincarnation because its quality improved stratospherically thanks to vdW encapsulation with hBN. Today, encapsulated graphene's charge carrier mobilities hold all materials' record of ~150,000 $cm^2$ $V^{-1}$ $s^{-1}$ at room temperature and reach several million at liquid-helium temperatures. Moreover, the latter mobilities are limited by edge scattering, that is, only by the maximum size (~10 μm) of currently available graphene devices. The high electronic quality has led to spectacular progress in various graphene subfields including spintronics, valleytronics, twistronics, electron fluidics, superlattices, etc. with the most recent example being the discovery of superconductivity and orbital ferromagnetism in twisted graphene multilayers. Hundreds if not thousands of reviews have already been dedicated to graphene, 2D materials and their vdW heterostructures, covering myriads of specialized topics.

The purpose of this Viewpoint is to provide a perspective on yet another emerging subfield, not reviewed so far. It developed out of nothing, both literally and figuratively. Remember how graphene is made? Take a piece of graphite, pull out an individual atomic plane and discard the bulk. Now switch your mental focus and forget about graphene: what would be left behind in graphite? It is an angstrom scale cavity, a pair of edge dislocations with an empty space in between. Such 2D cavities would be science fiction just a few years ago but, thanks to the vdW technology, the recipe to make them is conceptually simple: Prepare strips of graphene and use them as spacers between two atomically flat crystals as shown in Fig. 1. The trilayer vdW assembly allows a lot of flexibility in design of 2D cavities, using different vdW materials and spacers of different thicknesses.

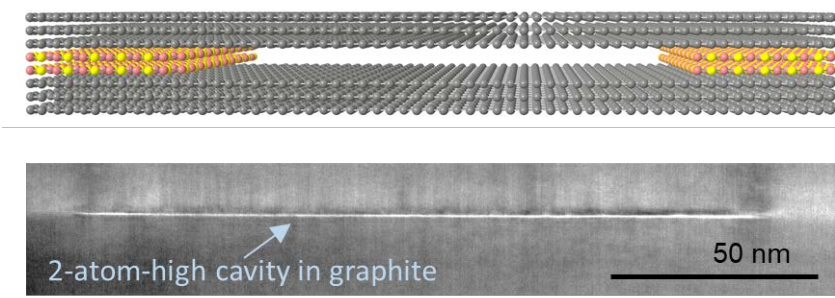

**Figure 1| 2D empty space**. (top) Artist's impression: angstrom-scale cavities can be designed to have different sizes and atomically flat walls made from different materials. (bottom) Transmission electron

micrograph of graphene's alter ego: 2D cavity in graphite with a nominal height of ~6.7 Å. Here, strips of bilayer graphene serve as spacers placed between two graphite crystals[3].

These are early days for research on 2D cavities, and their properties have been probed only in a few experiments. Nonetheless, the results already indicate a plethora of interesting science to come. For example, graphitic cavities of several nm in height – relatively large by 2D standards – exhibit exceptionally fast flows of water, with giant slip lengths exceeding 100 nm. Tighter cavities providing molecular-scale confinement (heights of < 2 nm) seem to enable even faster, nearly frictionless flows[3]. Not only water but also gases can flow through 2D cavities with little friction[4]. Helium atoms experience specular reflection from the inner walls of cavities made from graphene or hBN, which results in 'ballistic' gas transport, that is, a flow with no discernable friction. Although one might think that all atomically flat crystals are equally flat, some of them appear to be flatter than others. For example, 2D cavities made from $MoS_2$ do not exhibit specular reflection so that He atoms scatter randomly, resulting in conventional Knudsen behavior[4].

Even in hindsight, it is surprising that 2D cavities often display nearly ideal behavior. We deal here with physics determined by surfaces; essentially, it is surface science but without its usual attributes such as ultrahigh vacuum. 2D cavities are fabricated under ambient conditions where surfaces are covered with hydrocarbon contamination[4,5]. Some cavities do get blocked, but many remain clean so that their surfaces are apparently free or nearly free from adsorbates. It has been speculated that hydrocarbons get squeezed out of angstrom-scale 2D spaces, similar to self-cleansing that occurs during the standard vdW assembly in which 2D crystals are placed in direct contact, without spacers[5].

Of particular interest are 2D cavities with angstrom-scale heights because they provide sufficiently strong confinement for even the smallest molecular species to feel a squeeze. Using cavities with spacers made from bilayer graphene or monolayer $MoS_2$ (Fig. 1), ions in seawater can be sieved according to their hydrated diameters[6]. For example, a difference of nearly two orders of magnitude in mobility was observed between K and Al ions despite their hydrated diameters differing by less than 50%. The ultimately tight cavities have been made using monolayers of graphene or hBN as spacers. They admit a monolayer of water but reject all hydrated ions[7]. Only $H^+$ can permeate through this strictly 2D water expected to have a square-like structure. The proton transport was attributed to protons being able to jump between water molecules without the need of carrying around large hydration shells (Grotthuss mechanism)[7]. The monolayer cavities are the smallest manmade channels and their height of ~3.4 Å is close in size to aquaporins, short protein channels in biological cells. So far, 2D cavities cannot compete with protein channels in terms of their truly amazing selectivity for certain ions. Nonetheless, the artificial cavities are a leap forward for molecular transport experiments at extreme nanoscale, allowing to probe physics and phenomena that until recently were exclusively the domain of computer simulations.

Many problems that were previously inaccessible in experiment can now be investigated employing 2D cavities. One of them is capillary condensation. Under ambient humidity, water is known to spontaneously condense inside nanopores and between touching surfaces. The condensed liquid water strongly affects many commonplace phenomena such as friction, adhesion, corrosion, lubrication, stiction, etc. Under the normal humidity of 30-50%, condensation occurs for confinement of $\lesssim$ 2 nm, comparable to water molecule's diameter. Ludicrously, the equations currently used to describe the condensation are macroscopic, ignoring the discreteness of liquid water at this scale. This indicates a glaring gap in our understanding of ubiquitous condensation effects. The discussed 2D cavities provided a tool to accurately investigate this regime for the first time[8]. Although no curvature can possibly be assigned to menisci made of one or two monolayers of water, the century-old Kelvin equation was found to hold reasonably well for those ultimately tight capillaries, albeit fortuity was also involved in the observed agreement[8].

Another example of difficult problems that can be addressed using 2D cavities is dielectric response of interfacial water. Bulk water has exceptionally high polarizability with the dielectric constant $\varepsilon \approx 80$. Many phenomena including solvation, hydration and protein folding involve interfacial water, and the electrical double layer is ubiquitous in technology. As it stands, such nanoscale phenomena are routinely described using $\varepsilon$ of bulk water. This is unjustified, as the properties of water change under molecular-scale confinement. Using 2D cavities made from graphene and hBN, $\varepsilon$ was measured as a function of confinement[9]. Water's dielectric response is found to gradually decrease from its bulk value to $\varepsilon$ of only ~2 for nm-thick water. In effect, confined water becomes electrically dead, nonpolarizable in the direction perpendicular to the surfaces. It remains to be determined how water responds in the parallel direction. Nonetheless, it is clear that, if water's $\varepsilon$ is involved, analysis of those nanoscale phenomena requires extra care.

To conclude, although experiments using graphene's alter ego – 2D empty space – still remain the preserve of only a few groups around the world, they are expected to eventually become a useful tool for studying phenomena at the ultimately small scale of materials science, the size of individual atoms and molecules. Graphene is dead, long live graphene.